
  \magnification\magstep1
  \baselineskip = 0.5 true cm
                           
  \def\sa{\vskip 0.30 true cm}
  \def\sb{\vskip 0.60 true cm}
  \def\sc{\vskip 0.15 true cm}
  \def\sd{\vskip 0.50 true cm}
  \def\demi{ { {\lower 3 pt\hbox{$\scriptstyle 1$}} \over 
               {\raise 3 pt\hbox{$\scriptstyle 2$}} } } 

  \vsize = 25.7   true cm
  \hsize = 15.6   true cm
  \voffset = -1 true cm
  \parindent=0.85 true cm


  \rightline{\bf LYCEN~~9742}
  \rightline{September 1997} 

\sc
\sb
\sb 
\vskip 1 true cm  

\centerline {\bf THE $k$-FERMIONS AS OBJECTS INTERPOLATING} 

\vskip 0.8 true cm

\centerline {{\bf BETWEEN FERMIONS AND BOSONS}{\footnote{$^*$} 
{\sevenrm 
Work presented by one of the authors (M.K.) 
both to the Symposium ``Symmetries in Science X'' 
held at the Cloister Mehrerau (Bregenz, Austria, 13-18 July 1997) 
and to the ``VIII International Conference on Symmetry Methods in Physics'' 
held at the Joint Institute for Nuclear Research (Dubna, Russia, 28 July~-~2
August 1997). 
To be published in {\bf Symmetries in Science X}, 
eds.~B. Gruber and M. Ramek (Plenum Press, New York, 1998).}}}

\vskip 0.4 true cm                                       

\sa
\sb

\vskip 0.5 true cm

\centerline 
{M. Daoud,$^1$ Y. Hassouni$^1$ and M. Kibler$^2$}
\sa

\centerline {$^1$Laboratoire de Physique Th\'eorique}
\centerline {Universit\'e Mohammed V}
\centerline {Avenue Ibn Batouta, B.P. 1014}
\centerline {Rabat, Morocco}

\sa

\centerline {$^2$Institut de Physique Nucl\'eaire de Lyon}
\centerline {IN2P3-CNRS et Universit\'e Claude Bernard}
\centerline {43 Boulevard du 11 Novembre 1918}
\centerline {F-69622 Villeurbanne Cedex, France}

\sa

\sa
\sa
\sb
\sb

\centerline {\bf Abstract} 

\sa
\sa

Operators, refered to as $k$-fermion operators, that interpolate
between boson and  fermion  operators are introduced through the 
consideration of two noncommuting quon algebras. The deformation parameters 
$q$ and  $1/q$  for these quon algebras are roots of unity with 
$q = {\rm exp}( { 2 \pi {\rm i} / k } )$ and 
$k \in {\bf N} \setminus \{ 0,1 \}$. 
The case $k=2$ corresponds to fermions and
the limiting case $k \to \infty$ to bosons. 
Generalized coherent states
(connected to $k$-fermionic states) and supercoherent states 
(involving a  $k$-fermionic sector and a purely bosonic sector) are 
investigated. The operators in the $k$-fermionic algebra are used to find 
realisations of the Dirac quantum phase operator and of the $W_{\infty}$ 
Fairlie-Fletcher-Zachos algebra.

\baselineskip = 0.7 true cm

  \sa

\vfill\eject

\baselineskip = 0.5 true cm

  \vglue 4 true cm

\noindent {\bf THE $k$-FERMIONS AS OBJECTS INTERPOLATING} 

\sb

\noindent {\bf BETWEEN FERMIONS AND BOSONS} 

\vskip 0.4 true cm 

\sa
\sb

\vskip 0.5 true cm

\leftskip = 1.6 true cm

{M. Daoud,$^1$ Y. Hassouni$^1$ and M. Kibler$^2$} 

\sa 

{$^1$Laboratoire de Physique Th\'eorique}

{Universit\'e Mohammed V}

{Avenue Ibn Batouta, B.P. 1014}

{Rabat, Morocco}

\sa 

{$^2$Institut de Physique Nucl\'eaire de Lyon}

{IN2P3-CNRS et Universit\'e Claude Bernard}

{43 Boulevard du 11 Novembre 1918}

{F-69622 Villeurbanne Cedex, France}

\leftskip = 0 true cm

\sa
\sb
\sa
\sa
\sb
\sb

\baselineskip = 0.5 true cm

\noindent {\bf I. INTRODUCTION}

\sb

In the recent years, the theory of deformations, 
mainly in the spirit of quantum groups and quantum algebras, 
has been the subject of considerable interest in statistical physics. 
More precisely, deformed oscillator algebras have proved to be useful 
in {\it parastatistics} 
(connected to irreducible representations, 
of dimensions greater than 1, 
of the symmetric group), in {\it anyonic statistics} (connected to the braid
group) that concerns only particles in (one or) two dimensions, and in 
$q$-{\it deformed statistics} that may concern particles in arbitrary
dimensions. In particular, the $q$-deformed statistics deal with:  

\item{(i)} $q$-bosons    (which are bosons   obeying a $q$-deformed 
Bose-Einstein distribution),

\item{(ii)} $q$-fermions (which are fermions obeying a $q$-deformed 
Fermi-Dirac   distribution), and

\item{(iii)} quons (with $q$ such that $q^k = 1$, 
where $k \in {\bf N} \setminus \{ 0,1 \}$) which are objects, 
refered to as $k$-fermions in this work, interpolating between 
fermions (corresponding to $k=2$) and 
bosons   (corresponding to $k \to \infty$). 

This paper is devoted to $k$-fermions. A basic tool for studying such objects
is furnished by generalized Grassmann variables. These variables were
introduced in connection with quantum groups.$^{1-3}$ They constitute a natural
extension of ordinary Grassmann variables $z$, with $z^2 = 0$, which occur in
SUSY theories (basically, in the supersymmetric Poincar\'e group(s)). They play
an important role in fractional supersymmetry.$^{3 - 11}$  They arise in this 
work by looking for realizations of $k$-fermion operators involved in a quon
algebra$^{12-14}$ for which the deformation parameter $q$ is a root of unity.

The material in the present paper is organized as follows. The next section
(Section II) deals with $k$-fermions. We first discuss the quon algebras $A_q$
and $A_{\bar q}$, where $q := {\rm exp} (2 \pi {\rm i} / k)$, in terms of 
generalized Grassmann variables. Then, we introduce generalized coherent 
states. Finally, the notion of a fractional supercoherent state is introduced 
from a certain limit
of the well-known (see Refs.~12 and 15 to 20) $q$-deformed coherent states. 
Section III is devoted to the quantum phase operator. The last section 
(Section IV) is concerned with symmetries, 
described by the algebras $W_{\infty}$ and $U_q(sl(2))$, inherent to 
the introduction of $k$-fermions. 

\sb
\sd
  

\noindent {\bf II. INTRODUCING $k$-FERMIONS} 

\sb

\noindent {\bf 1. The quon algebras $A_q$ and $A_{\bar q}$} 

\sa

Let us first start with the quon algebra $A_q$. The algebra 
$A_q$ is generated by an annihilation operator $a_-$, 
                           a creation operator $a_+$ and a number operator $N$ 
with the relations 
$$
a_- a_+ - q a_+ a_- = 1
\eqno (1)
$$
and 
$$
N a_- - a_- N = - a_-
\eqno (2{\rm a})
$$
$$
N a_+ - a_+ N = + a_+
\eqno (2{\rm b})
$$
Here and in the following, the complex number $q$ is chosen to be 
$$
q := {\rm exp} \left( {2 \pi {\rm i} \over k} \right) 
\eqno (3)
$$
where $k$ is a fixed number in ${\bf N} \setminus \{ 0,1 \}$. 
In other words, $q$ is a root of unity. Furthermore, 
the operator $N$ is taken to be hermitean. Equation (1), with $q$ being a root
of unity, shows that the operator $a_-$ (respectively, $a_+$) 
cannot be (except for $k = 2$ and $k \to \infty$) 
the adjoint of       the operator $a_+$ (respectively, $a_-$). Note that
Eq.~(1) is satisfied by 
$$
a_- a_+ = \left[ N + s + {1 \over 2} \right]_q
\eqno (4{\rm a})
$$
$$
a_+ a_- = \left[ N + s - {1 \over 2} \right]_q
\eqno (4{\rm b})
$$
where $s = {1 \over 2}$ and the symbol $[~~]_q$ is defined by (see the appendix)
$$
[X]_q := {1 - q^X \over 1 - q}
\eqno (5)
$$
where $X$ may be an operator or a number. 

From Eq.~(1), we obtain
$$
a_- (a_+)^{\ell} = [\ell]_q (a_+)^{\ell-1} + q^{\ell} (a_+)^{\ell} a_-
\eqno (6{\rm a})
$$
and
$$
(a_-)^{\ell} a_+ = [\ell]_q (a_-)^{\ell-1} + q^{\ell} a_+ (a_-)^{\ell}
\eqno (6{\rm b})
$$
for $\ell = 1, 2, \cdots, k-1$. In the particular case $\ell =k$, Eqs.~(6a) and
(6b) are amenable to the form 
$$
a_- (a_+)^k = (a_+)^k a_-
\eqno (7{\rm a})
$$
and
$$
(a_-)^k a_+ = a_+ (a_-)^k
\eqno (7{\rm b})
$$
In addition, Eqs.~(2a) and (2b) yield 
$$
N (a_+)^{\ell} = (a_+)^{\ell} (N + \ell)
\eqno (8{\rm a})
$$
and
$$
(a_-)^{\ell} N = (N + \ell) (a_-)^{\ell}
\eqno (8{\rm b})
$$
for $\ell = 1, 2, \cdots, k$. Equations (8) with $\ell = k$ 
and  (7)  are nothing but trivial identities if we assume 
$$
(a_+)^k = 0
\eqno (9{\rm a})
$$
$$
(a_-)^k = 0
\eqno (9{\rm b})
$$
In this paper, we shall deal with a representation of the algebra 
$A_q$ such that Eqs.~(9a) and (9b) are satisfied. Remark that, should we have
defined $A_q$ by Eq.~(1) only, we would have two further types of
representations, viz.~the periodic representation for which
$$
(a_+)^k = 1 \quad {\hbox{and}} \quad (a_-)^k = 1
\eqno (10)
$$
and the semi-periodic representations for which
$$
(a_+)^k = 1 \quad {\hbox{and}} \quad (a_-)^k = 0
\eqno (11{\rm a})
$$
or
$$
(a_+)^k = 0 \quad {\hbox{and}} \quad (a_-)^k = 1
\eqno (11{\rm b})
$$

Going back to the  representation  for which 
the nilpotency conditions (9) are fulfilled, 
we note that the algebra $A_{-1}$, obtained for $k=2$, corresponds to ordinary 
fermion operators with $(a_-)^2 = (a_+)^2 = 0$, 
a relation that reflects the Pauli exclusion principle. In the limiting case 
$k \to \infty$, we have the algebra $A_{+1}$ which  corresponds to ordinary  
boson   operators~; the algebra $A_{+1}$ can thus be considered as the 
oscillator algebra well-known in quantum mechanics. For $k$ arbitrary, the 
algebra $A_q$ corresponds to quon operators (or $k$-fermion operators) 
$a_-$  and  $a_+$ that interpolate between fermion and boson operators. 

We continue with the situation where $k \in {\bf N} \setminus \{ 0,1 \}$ 
and where the constraints (9) hold. In this situation, we easily obtain the 
$k$-dimensional representation of $A_q$ defined through
$$
a_- | n   \rangle = \left( \left[ n + s - {1 \over 2} \right]_q \right)^{{1 \over2}} 
    | n-1 \rangle \quad {\hbox{with}} \quad a_- |   0 \rangle = 0
\eqno (12{\rm a})
$$
$$
a_+ | n   \rangle = \left( \left[ n + s + {1 \over 2} \right]_q \right)^{{1 \over2}} 
    | n+1 \rangle \quad {\hbox{with}} \quad a_+ | k - 1 \rangle = 0
\eqno (12{\rm b})
$$
and
$$
N | n \rangle = n | n \rangle 
\eqno (13)
$$
where $n = 0, 1, \cdots, k-1$. This representation is 
builded on a finite-dimensional (Fock) unitary space 
${\cal F} := \{ | n \rangle : n = 0, 1, \cdots, k-1 \} $
of dimension $k$. The state vector $ | n \rangle $
is given by 
$$
| n \rangle = { (a_+)^n \over ([n]_q !)^{1 \over 2} } \> |0 \rangle \quad
{\hbox{for}} \quad n = 0, 1, \cdots, k-1
\eqno (14)
$$
where, as usual, the $[n]_q$-factorial is defined by 
$$
\lbrack n \rbrack_q ! := 
\lbrack 1 \rbrack_q 
\lbrack 2 \rbrack_q \cdots 
\lbrack n \rbrack_q \quad {\hbox{for}} \quad 
n \in {\bf N} \setminus \lbrace 0 \rbrace \qquad {\hbox{and}} \qquad
\lbrack 0 \rbrack_q ! := 1
\eqno (15)
$$
The space ${\cal F}$ is of dimension 2 for the fermionic algebra $A_{-1}$
and infinite-dimensional               for the bosonic   algebra $A_{+1}$. 

We now come to the algebra $A_{\bar q}$. 
Let $a_+^+$ (respectively, $a_-^+$) be the adjoint operator of 
    $a_+$   (respectively, $a_-$). The operators $a_+^+$, 
                                                 $a_-^+$ and $N$ span the quon
algebra $A_{\bar q}$ since 
$$
a_+^+ a_-^+ - {\bar q} a_-^+ a_+^+ = 1
\eqno (16)
$$
and
$$
N a_+^+ - a_+^+ N = - a_+^+
\eqno (17{\rm a})
$$
$$
N a_-^+ - a_-^+ N = + a_-^+
\eqno (17{\rm b})
$$
where 
$$
{\bar q} = {\rm exp} \left( - {2 \pi {\rm i} \over k} \right)
\eqno (18)
$$
is the complex conjugate of $q$. Thus, $a_+^+$ plays the role of an 
annihilation operator and              $a_-^+$ the one of a creation operator
for the algebra $A_{\bar q}$. 
A representation similar to the one described for $A_q$ may be obtained for 
$A_{\bar q}$. This $k$-dimensional representation corresponds to Eq.~(13) 
supplemented with 
$$
a_+^+ | n \rangle = 
\left( \left[ n + s - {1 \over 2} \right]_{\bar q} \right)^{1 \over 2} | n-1 \rangle
\quad {\hbox{with}} \quad a_+^+ |   0 \rangle = 0
\eqno (19{\rm a})
$$
$$
a_-^+ | n \rangle = 
\left( \left[ n + s + {1 \over 2} \right]_{\bar q} \right)^{1 \over 2} | n+1 \rangle
\quad {\hbox{with}} \quad a_-^+ | k-1 \rangle = 0
\eqno (19{\rm b})
$$

At this point, it is worth noticing that the relation 
$$
a_- a_+^+ = q^{-{1 \over 2}} a_+^+ a_- \iff 
a_+ a_-^+ = q^{+{1 \over 2}} a_-^+ a_+
\eqno (20)
$$
holds, on the space ${\cal F}$, for the representations 
of $A_q$ and $A_{\bar q}$ under consideration. 

\sb

\noindent {\bf 2. Towards Grassmannian realizations of $A_q$ and $A_{\bar q}$} 

\sa

We are now in a position to deal with realizations of the algebras $A_q$ 
and $A_{\bar q}$ in terms of generalized Grassmann variables. Equations 
(9) suggest that we use generalized Grassmann variables 
(see Refs.~1 to 3 and 21) $z$ and ${\bar z}$ 
such that  
$$
z^k = 0
\eqno (21)
$$
and
$$
{\bar z}^k = 0
\eqno (22)
$$
where $k$ is a fixed number in ${\bf N} \setminus \{ 0,1 \}$. 
(The particular case $k = 2$ corresponds to ordinary Grassmann variables.) We 
then introduce the $\partial_{z}$- and $\partial_{\bar z}$-derivatives via 
$$
\partial_z f(z) := {f(qz) - f(z) \over (q - 1) z}
\eqno (23)
$$
and
$$
\partial_{\bar z} g({\bar z}) := 
{g({\bar q} {\bar z}) - g({\bar z}) \over ({\bar q} - 1) {\bar z}}
\eqno (24)
$$
where $f$ and $g$ are arbitrary functions of $z$ and $\bar z$, respectively. 
The linear operators $\partial_{z}$ and $\partial_{\bar z}$ satisfy 
$$
\partial_z z^n = [n]_q \> z^{n-1}
\eqno (25)
$$
and
$$
\partial_{\bar z} {\bar z}^n = [n]_{\bar q} \> {\bar z}^{n-1}
\eqno (26)
$$
for $n = 0, 1, \cdots, k-1$. Therefore, for functions 
$f :       z  \mapsto f(      z )$ and 
$g : {\bar z} \mapsto g({\bar z})$ that can be developed as
$$
f(z) = \sum_{n=0}^{k-1} a_n z^n
\eqno (27)
$$
and
$$
g({\bar z}) = \sum_{n=0}^{k-1} b_n {\bar z}^n
\eqno (28)
$$
where the coefficients $a_n$ and $b_n$ in the expansions are complex numbers,
we check that 
$$
(\partial_z)^k f(z) = 0
\eqno (29)
$$
and
$$
(\partial_{\bar z})^k g({\bar z}) = 0
\eqno (30)
$$
Consequently, we shall assume that the conditions 
$$
(\partial_z)^k = 0
\eqno (31)
$$
and
$$
(\partial_{\bar z})^k = 0
\eqno (32)
$$
hold in addition to Eqs.~(21) and (22). 

The correspondences 
$$
a_-   \rightarrow \partial_z        \qquad a_+   \rightarrow       z
\eqno (33)
$$
and
$$
a_+^+ \rightarrow \partial_{\bar z} \qquad a_-^+ \rightarrow {\bar z}
\eqno (34)
$$
clearly provide us with a realization of Eqs.~(1) and (16). 
In this realization, Eqs.~(21) and (31) are the images of   
Eqs.~(9a) and (9b), respectively. Similarly, Eqs.~(22) and 
(32) are the images of the adjoint relations of Eqs.~(9b) 
and (9a), respectively. Note that Eq.~(20) leads to
$$
z {\bar z} = q^{+{1 \over 2}} \> {\bar z} z
\eqno (35)
$$
and
$$
\partial_z \partial_{\bar z} = q^{-{1 \over 2}} \> \partial_{\bar z} \partial_z
\eqno (36)
$$
in the realization based on Eqs.~(33) and (34).

\sb

\noindent {\bf 3. Generalized coherent states} 

\sa 

There exists several methods for introducing coherent states. We can use the
action of a displacement operator on a reference state$^{22}$
or the construction of an eigenstate for an annihilation operator$^{23,24}$
or the minimisation of uncertainty relations.$^{25}$ In the case of the 
ordinary harmonic oscillator, the three methods lead to the same result (when
the reference state is the vacuum state). Here, the situation is a little bit
more intricate (as far as the equivalence of the three methods is concerned) 
and we chose to define the generalized 
                           coherent states or $k$-fermionic coherent states 
$|       z )$ and 
$| {\bar z})$ as follows
$$
|z)        := \sum_{n=0}^{k-1} {       z ^n \over ([n]_q       !)^{1 \over 2} } 
\> | n \rangle
\eqno (37)
$$ 
and 
$$
|{\bar z}) := \sum_{n=0}^{k-1} { {\bar z}^n \over ([n]_{\bar q}!)^{1 \over 2} }
\> | n \rangle
\eqno (38)
$$ 
where $z$ and $\bar z$ are generalized Grassmann variables that satisfy
Eqs.~(21) and (22). It can be easily checked that 
the state vectors $| z )$ and $| {\bar z} )$ are eigenvectors of the operators
$a_-$ and $a_+^+$, respectively. More precisely, we have
$$
a_- |z) = z | z)
\eqno (39)
$$
and 
$$
a_+^+ | {\bar z} ) = {\bar z} | {\bar z} )
\eqno (40)
$$
The case $k=2$ corresponds to fermionic coherent states while the limiting case
$k \to \infty$ to bosonic coherent states.

We define 
$$
( z | := \sum_{n=0}^{k-1} \> \langle n | \>
{ {\bar z}^n \over ([n]_{\bar q}!)^{1 \over 2} }
\eqno (41)
$$
and 
$$
( {\bar z} | := \sum_{n=0}^{k-1} \> \langle n | \>
{ z^n \over ([n]_q!)^{1 \over 2} } 
\eqno (42)
$$
Then, the `scalar products' $( z' | z )$ and $( {\bar z}' | {\bar z} )$ 
follow  from  the ordinary scalar product 
$ \langle n' | n \rangle = \delta(n',n)$. For instance, we get 
$$
(z' | z) = \sum_{n=0}^{k-1} 
              { {\bar {z'}}^n z^n \over ([n]_{\bar q}! [n]_q!)^{1 \over 2} }
\eqno (43)
$$
In view of the relationship 
$$
[n]_{\bar q}!  = q^{- {1 \over 2} n(n-1)} \> [n]_q !
\eqno (44)
$$
and of the property (cf.~Eq.~(35))
$$
{\bar z}^n z^n = q^{- {1 \over 4} n(n-1)} \> ({\bar z} z)^n
\eqno (45)
$$
we obtain the following result 
$$
(z | z) = \sum_{n=0}^{k-1} { ({\bar z} z)^n \over [n]_q! }
\eqno (46)
$$
Similarly, we have 
$$
({\bar z} | {\bar z}) = \sum_{n=0}^{k-1} { (z {\bar z})^n \over [n]_{\bar q}! }
\eqno (47)
$$
By defining the $q$-deformed exponential ${\rm e}_q$ by 
$$
{\rm e}_q : x \mapsto {\rm e}_q (x) := \sum_{n=0}^{k-1} { x^n \over [n]_q! }
\eqno (48)
$$
we can rewrite Eqs.~(46) and (47) as 
$$
(       z |      z  ) = {\rm e}_      q  ({\bar z} z) \quad {\hbox{and}} \quad 
( {\bar z}|{\bar z} ) = {\rm e}_{\bar q} (z {\bar z})
\eqno (49)
$$
(Observe that the summation 
in the exponential ${\rm e}_q$ is finite, rather than 
infinite as is usually the case in $q$-deformed exponentials.)

We guess that the $k$-fermionic coherent states $| z )$ and $| {\bar z} )$ 
form overcomplete sets with respect to some integration process 
accompanying the derivation process inherent to Eqs.~(23) and (24). 
Following Majid and Rodr\'\i guez-Plaza,$^{21}$ we consider the integration
process defined by
$$
\int dz \> z^p     = \int d{\bar z} \> {\bar z}^p := 0 \quad {\hbox{for}} 
\quad p = 0, 1, \cdots, k-2
\eqno (50{\rm a})
$$
and
$$
\int dz \> z^{k-1} = \int d{\bar z} \> {\bar z}^{k-1} := 1
\eqno (50{\rm b})
$$
Clearly, the integrals in (50) generalize the Berezin integrals corresponding
to $k = 2$. In the case where $k$ is arbitrary, we can derive the
overcompleteness property 
$$
\int d      z  \> |       z  ) \> 
\mu(z, {\bar z}) \> (       z  | \> d{\bar z} = 
\int d{\bar z} \> | {\bar z} ) \> 
\mu({\bar z}, z) \> ( {\bar z} | \> d      z  = 1
\eqno (51)
$$
where the function $\mu$ defined through
$$
\mu(z, {\bar z}) := \sum_{n=0}^{k-1} \> ( [n_q]! [n_{\bar q}]! )^{1 \over 2} \>
z^{k-1-n} \> {\bar z}^{k-1-n}
\eqno (52)
$$
may be regarded as a measure.

\sb

\noindent {\bf 4. The coherence factor} 

\sa

The coherence factor turns out to be of central importance in 
quantum optics and multiphoton spectroscopy.  It occurs in the 
expression of the absorption or emission multiphoton intensity. 
We define here the coherence factor $g^{(m)}$ of order $m$ 
(with $ m \in {\bf N} \setminus \{ 0 \} $) by
$$
g^{(m)} := { \langle \left( a_-^+ \right)^m 
                     \left( a_-   \right)^m \rangle \over 
             \langle a_-^+ a_- \rangle^m }
\eqno (53)
$$
where we have adopted the notation
$$
\langle X \rangle := { (z | X | z) \over (z | z) }
\eqno (54)
$$
for denoting the average value of the operator $X$ on the state vector $|z)$. 
It is then a simple matter of calculation to show that
$$
g^{(m)} = { {\bar z}^m z^m \over ({\bar z} z)^m } \> \theta (k - 1 - m)
\eqno (55)
$$
where the step function $\theta$ is such that
$$
\theta (t) := 0 \quad {\hbox{if}} \quad t < 0 \qquad {\hbox{and}} \qquad
\theta (t) := 1 \quad {\hbox{if}} \quad t \geq 0
\eqno (56)
$$
By introducing (45) into (55), we end up with the simple formula
$$
g^{(m)} = q^{- {1 \over 4} m (m-1)} \> \theta (k-1-m)
\eqno (57)
$$
This formula generalizes well-known results for fermions and bosons. Indeed,
we have
$$
g^{(1)} = 1 \qquad {\hbox{and}} 
            \qquad g^{(m)} = 0 \quad {\hbox{for}} \quad m \geq 2
\eqno (58)
$$
for fermions (corresponding to $k=2$) and
$$
g^{(m)} = 1 \quad {\hbox{for}} \quad m \geq 1
\eqno (59)
$$
for bosons (corresponding to $k \to \infty$). In the case where $k$ is 
arbitrary, the vanishing of the coherence factor $g^{(m)}$ for $m > k-1$
is an indication, in a many-particle scheme, that a given quantum state 
(of fractional spin $S = {1 \over k}$) cannot be
occupied by more than $k-1$ identical particles. 
We shall refer the quons described 
by the couple ($A_q, A_{\bar q}$) to as $k$-fermions. 
A field constituted of $k$-fermions thus satisfies a generalized Pauli
exclusion principle. The case $k=2$ corresponds to ordinary fermions 
(which obey the Pauli exclusion principle)  
while the case $k \to \infty$ corresponds to ordinary
bosons (which do not obey the Pauli exclusion principle). 

\sb

\noindent {\bf 5. Fractional supercoherent states} 

\sa

We now switch to $Q$-deformed coherent states of the type 
$$
|Z) := \sum_{n=0}^{\infty} { Z^n \over ([n]_Q !)^{1 \over 2} } \> | n \rangle
\eqno (60)
$$
associated to a quon algebra $A_Q$ where $Q \in {\bf C} \setminus {S}^1$.
The latter states are simple deformations of the bosonic coherent states
(cf.~Ref.~12). The coherent state $| Z )$ may be considered 
to be an eigenstate, with the eigenvalue $Z \in {\bf C}$, of an annihilation
operator $b_-$ in a representation such that the operator $b_-$ and 
the associated creation operator $b_+$ satisfy
$$
b_- | n \rangle = \left( \left[ n + s - {1 \over 2} \right]_Q \right)^{1 \over 2} 
| n - 1 \rangle \quad {\hbox{with}} \quad b_- | 0 \rangle = 0
\eqno (61{\rm a})
$$
$$
b_+ | n \rangle = \left( \left[ n + s + {1 \over 2} \right]_Q \right)^{1 \over 2} 
| n + 1 \rangle
\eqno (61{\rm b})
$$
with $n \in {\bf N}$.  

For $Q \to q$, we have $[k]_Q! \to 0$. Therefore, the term 
$Z^k / ( [k]_Q! )^{1 \over 2}$ 
in Eq.~(60) makes sense for $Q \to q$ only if $Z \to z$,
where $z$ is a generalized 
Grassmann variable with $z^k = 0$. This type of reasoning has
been invoked for the first time in Refs.~26 and 27. (In these references,
the authors show that there is an isomorphism between the braided line and the
one-dimensional superspace.)

It is the aim of this section to determine the limit 
$$
| \xi ) := \lim_{Q \to q} \lim_{Z \to z} \> |Z)
\eqno (62)
$$
when $Q$ goes to the root of unity $q$ (see Eq.~(3)) and $Z$ to a Grassmann
variable $z$. The starting point is to rewrite Eq.~(60) as 
$$
|Z) = \sum_{r=0}^{\infty} \sum_{s=0}^{k-1} 
{ Z^{rk + s} \over ( [rk + s]_Q! )^{1 \over 2} } \> | rk + s )
\eqno (63)
$$
Then, by making use of the formulas
$$
{ [k]_Q \over [r k]_Q } \to {1 \over r} \quad {\hbox{for}} \quad 
Q \to q \quad {\hbox{with}} \quad r \ne 0
\eqno (64{\rm a})
$$
and
$$
{ [s]_Q \over [rk + s]_Q } \to 1 \quad {\hbox{for}} \quad 
Q \to q \quad {\hbox{with}} \quad s = 0, 1, \cdots, k-1
\eqno (64{\rm b})
$$
we find that 
$$
\lim_{Q \to q} \lim_{Z \to z} 
{ Z^{rk + s} \over \left( [rk + s]_Q ! \right)^{1 \over 2} } 
= { z^s      \over ([s_q]!)^{1 \over 2} } \> 
  { \alpha^r \over (r    !)^{1 \over 2} }
\eqno (65)
$$
works for $s = 0, 1, \cdots, k-1$ and $r \in {\bf N}$. 
The complex variable $\alpha$ in Eq.~(65) is defined by 
$$
\alpha := 
\lim_{Q \to q} \lim_{Z \to z} { Z^k \over \left( [k]_Q! \right)^{1 \over 2} }
\eqno (66)
$$
Therefore, we obtain
$$
| \xi ) = \sum_{r=0}^{\infty} \sum_{s=0}^{k-1} 
{ z^s      \over \left( [s]_q! \right)^{1 \over 2} } 
{ \alpha^r \over (r!)^{1 \over 2} } \> | rk+s \rangle
\eqno (67)
$$
Finally, by employing the symbolic notation
$$
| rk + s \rangle \equiv | r \rangle \otimes | s \rangle
\eqno (68)
$$
we arrive at the formal expression 
$$
| \xi ) = 
\sum_{r=0}^{\infty} { \alpha^r \over (r!)^{1 \over 2} } \> | r \rangle 
\bigotimes 
\sum_{s=0}^{k-1}    { z^s \over \left( [s]_q! \right)^{1 \over 2} } \> 
                                                           | s \rangle
\eqno (69)
$$

We thus end up with the product of 
a       bosonic coherent state by 
a $k$-fermionic coherent state. This product shall be called a fractional 
supercoherent state. In the particular case $k=2$, it reduces to the product
of a    bosonic coherent  state by 
   a  fermionic coherent  state, i.e., to the supercoherent state associated to 
a superoscillator.$^{28}$ For $k$ arbitrary, 
the fractional supercoherent states are presently under study.$^{29}$ In the
framework of field theory, Eq.~(69) means that in the limit $Q \to q$, every
field $\psi$ with values $\psi(Z)$ is transformed into a fractional superfield
$\Psi$ with value $\Psi(z, \alpha)$, $z$ being a generalized Grassmann variable
and $\alpha$ a bosonic variable. 

\sb 
\sd

\noindent {\bf III. THE QUANTUM PHASE OPERATOR}
                    
\sb

The quantization of the phase angle of an harmonic oscillator was achieved by
Dirac$^{30}$ (see the review in Ref.~31). Pegg and Barnett$^{32}$ introduced a
phase operator in a finite-dimensional Hilbert space. In this direction,
Ellinas$^{33}$ used a troncated infinite-dimensional Hilbert space via the
consideration of semi-periodic representations of the quon algebra. We first 
briefly recall some classical results before giving an original approach 
to the (quantum) phase operator based on the use of $A_q$ and $A_{\bar q}$. 

\sb

\noindent {\bf 1. The classical approach} 

\sa

Let us define the phase states $| \theta_m )$ in the $k$-dimensional space
${\cal F}$ by 
$$
|\theta_m) := { 1 \over \sqrt{k} } \sum_{n=0}^{k-1} 
{\rm exp} ({\rm i} n \theta_m) \> | n \rangle \qquad m = 0, 1, \cdots, k-1
\eqno (70)
$$
where 
$$
\theta_m := \theta_0 + 2 \pi {m \over k}
\eqno (71)
$$
In Eq.~(71), $\theta_0$ is an arbitrary (real) reference angle. We have 
$( \theta_{m'} | \theta_{m} ) = \delta(m',m)$ so that the passage from the 
$| n \rangle $ basis to the $| \theta_m )$ basis (and vice versa) corresponds 
to a unitary transformation. The $n$--$m$ matrix element of this transformation 
is
$$
{ 1 \over \sqrt{k} } \> {\rm exp} ({\rm i} n \theta_m) \equiv 
{ 1 \over \sqrt{k} } \> {\rm exp} ({\rm i} n \theta_0) \> q^{mn}
\eqno (72)
$$
Then, the inverse of Eq.~(70) reads 
$$
| n \rangle = { 1 \over \sqrt{k} } \sum_{m=0}^{k-1} 
{\rm exp} ( - {\rm i} n \theta_m ) \> | \theta_m ) \qquad n=0, 1, \cdots, k-1
\eqno (73)
$$

Let us now introduce the phase operator $\phi$ as 
$$
\phi := \sum^{k-1}_{m=0} \theta_m \> | \theta_m ) ( \theta_m |
\eqno (74)
$$
This operator was introduced by Pegg and Barnett.$^{32}$ 
It is clear that $ | \theta_m ) $ is an eigenvector of 
$\phi$ with the real eigenvalue $\theta_m$. We now define the operators 
${\rm e}^{ + {\rm i} \phi}$ and
${\rm e}^{ - {\rm i} \phi}$ in the usual way, namely
$$
{\rm e}^{ \pm {\rm i} \phi } := \sum_{r=0}^{\infty} 
{ {(\pm {\rm i} \phi)^r} \over {r!} }
\eqno (75)
$$
The action of the latter operators on the space ${\cal F}$ is easily found 
to be characterized by
$$
{\rm e}^{ + {\rm i} \phi } | n   \rangle = | n-1 \rangle \quad {\hbox{for}} 
\quad n \ne 0   \qquad {\hbox{and}} \qquad 
{\rm e}^{ + {\rm i} \phi } | 0   \rangle =        \omega_{+k}   | k-1 \rangle
\eqno (76)
$$
and 
$$
{\rm e}^{ - {\rm i} \phi } | n   \rangle = | n+1 \rangle \quad {\hbox{for}} 
\quad n \ne k-1 \qquad {\hbox{and}} \qquad 
{\rm e}^{ - {\rm i} \phi } | k-1 \rangle =        \omega_{-k}   |   0 \rangle
\eqno (77)
$$
where
$$
\omega_{\pm k} := {\rm exp} ( \pm {\rm i} k \theta_0 )
\eqno (78)
$$
The matrix 
    representatives $M({\rm e}^{ + {\rm i} \phi})$ and   
                    $M({\rm e}^{ - {\rm i} \phi})$ of the operators
${\rm e}^{ + {\rm i} \phi}$ and
${\rm e}^{ - {\rm i} \phi}$ in the representation inherent to Eqs.~(76) and 
(77) are 
$$
M({\rm e}^{+ {\rm i} \phi}) =        \omega_{+k}   E_{k-1,0} 
+ \sum_{i=1}^{k-1} E_{i-1,i}
\eqno (79)
$$
and
$$
M({\rm e}^{- {\rm i} \phi}) =        \omega_{-k}   E_{0,k-1} 
+ \sum_{i=1}^{k-1} E_{i,i-1}
\eqno (80)
$$
where the matrices $E_{\alpha,\beta}$ are generators of the group SU($k$). 
As an interesting property, it is straightforward to derive 
$$
( {\rm e}^{\pm {\rm i} \phi} )^k = \omega_{ \pm k } 
\eqno (81)
$$
(The periodicity relations (81) trivially follow from the matrix representatives
(79) and (80).)

\sb

\noindent {\bf 2. A quon approach} 

\sa

An appealing question is to ask whether it is possible to define operators,
in terms of the generators of $A_{q}$ and $A_{\bar q}$, which exhibit a
property similar to the property (81). A positive answer follows from the
definition of the polynomials
$$
{\rm E}^{+ {\rm i} \Phi} := \left( [k - 1]_q!        \right)^{ -{1 \over k} } 
\left[ a_-   + \omega_{+k}          (a_+)  ^{k-1} \right]
\eqno (82)
$$
and  
$$
{\rm E}^{- {\rm i} \Phi} := \left( [k - 1]_{\bar q}! \right)^{ -{1 \over k} } 
\left[ a_-^+ + \omega_{-k}          (a_+^+)^{k-1} \right]
\eqno (83)
$$
where we use a notation that parallels the one for 
${\rm e}^{ + {\rm i} \phi}$ and
${\rm e}^{ - {\rm i} \phi}$. The action of the operators  
${\rm E}^{ + {\rm i} \Phi}$ and
${\rm E}^{ - {\rm i} \Phi}$ on the space ${\cal F}$ can be 
summarized as
$$
{\rm E}^{ + {\rm i} \Phi } | n   \rangle = 
\left( [ k - 1 ]_q!        \right)^{- {1 \over k}} 
\left( [n]_q        \right)^{1 \over 2}
| n-1 \rangle \quad {\hbox{for}} 
\quad n \ne 0   
$$
$$ 
{\rm E}^{ + {\rm i} \Phi } | 0   \rangle = 
\left( [ k - 1 ]_q!        \right)^{{1 \over 2} - {1 \over k}}
       \omega_{+k}   | k-1 \rangle
\eqno (84)
$$
and 
$$
{\rm E}^{ - {\rm i} \Phi } | n   \rangle = 
\left( [ k - 1 ]_{\bar q}! \right)^{- {1 \over k}} 
\left( [n]_{\bar q} \right)^{1 \over 2}
| n+1 \rangle \quad {\hbox{for}} 
\quad n \ne k-1 
$$
$$
{\rm E}^{ - {\rm i} \Phi } | k-1 \rangle = 
\left( [ k - 1 ]_{\bar q}! \right)^{{1 \over 2} - {1 \over k}}
       \omega_{-k}   |   0 \rangle
\eqno (85)
$$
It is to be remarked that the products
${\rm E}^{ + {\rm i} \Phi} \> {\rm E}^{ - {\rm i} \Phi}$
and 
${\rm E}^{ - {\rm i} \Phi} \> {\rm E}^{ + {\rm i} \Phi}$ 
are diagonal in the representations afforded by Eqs.~(79) and 
(80). Finally, we can prove that
$$
( {\rm E}^{\pm {\rm i} \Phi} )^k = \omega_{ \pm k } 
\eqno (86)
$$
to be compared with Eq.~(81). Note however that a true definition of the
quantum phase operator $\Phi$ requires that we give a sense to the
`exponentiation' ${\rm E}^{\pm {\rm i} \Phi}$. 

\sb 
\sd

\noindent {\bf IV. THE $W_{\infty}$ SYMMETRY}

\sb

We now introduce two couples of operators, ($U,V$) and ($X,Y$), defined in terms
of the generators of the algebras $A_{q}$ and $A_{\bar q}$, respectively. Let 
$$
V \equiv {\rm E}^{+ {\rm i} \Phi} \qquad U := a_-   a_+   - a_+   a_-  
\eqno (87)
$$
and
$$
Y \equiv {\rm E}^{- {\rm i} \Phi} \qquad X := a_+^+ a_-^+ - a_-^+ a_+^+ 
\eqno (88)
$$
where ${\rm E}^{\pm {\rm i} \Phi}$ are given by Eqs.~(82) and (83). 
Obviously, the couples ($U,V$) and ($X,Y$) are connected via hermitean
conjugation.

We first deal with the couple ($U,V$). An elementary calculation leads to 
$$
VU = q UV
\eqno (89)
$$
The latter relation can be iterated to produce
$$
V^n U^m = q^{nm} U^m V^n \qquad (n,m) \in {\bf N}^2
\eqno (90)
$$
Let us define the operator
$$
T_{(n_1,n_2)} := q^{ {1 \over 2} n_1n_2 } U^{n_1} V^{n_2} \qquad 
                 (n_1,n_2) \in {\bf N}^2
\eqno (91)
$$
It is convenient to use the abbreviation $(n_1, n_2) \equiv n$ so that
$$
T_{n} \equiv T_{(n_1,n_2)} 
\eqno (92)
$$
The product $T_nT_m$ is easily obtained to be
$$
T_m T_n = q^{ - {1 \over 2} m \times n } T_{m + n}
\eqno (93)
$$
where
$$
m \times n := m_1 n_2 - m_2 n_1 \qquad m + n = (m_1 + n_1, m_2 + n_2) 
\eqno (94)
$$
The commutator $[T_m,T_n] := T_m T_n - T_n T_m$ follows from Eq.~(93). 
Indeed, we have
$$
[T_m,T_n] = - 2 {\rm i} \sin \left( { \pi \over k } m \times n \right) T_{m + n}
\eqno (95)
$$
As a conclusion, the operators $T_m$ can be viewed as the generators of the
infinite dimensional Lie algebra $W_{\infty}$ (or sine algebra) 
investigated by 
Fairlie, Fletcher and Zachos.$^{34}$ A similar result can be derived from the
couple ($X, Y$). The occurrence of the algebra $W_{\infty}$ as a symmetry 
algebra for a system of $k$-fermions is still an open problem. 

To close this paper, we note that we can use the construction by Sato$^{35}$ 
and Kogan$^{36}$ for generating the quantum universal enveloping algebra
$U_q(sl(2))$. As a matter of fact, by defining the operators $J_+$, $J_-$ and 
$J_3$ through
$$
J_+ := { {T_{( 1, 1)} - T_{(-1,1)}} \over {q - q^{-1}} } 
\eqno (96)
$$
$$
J_- := { {T_{(-1,-1)} - T_{(1,-1)}} \over {q - q^{-1}} } 
\eqno (97)
$$
and
$$
q^{   2 J_3} := T_{ (- 2,0) } \qquad 
q^{ - 2 J_3} := T_{ (  2,0) }
\eqno (98)
$$
we can show that
$$
[J_+ , J_-] = { {q^{2J_3} - q^{-2J_3}} \over {q - q^{-1}} }
\eqno (99)
$$
and
$$
q^{2J_3} J_{\pm} q^{-2J_3} = q^{\pm 2} J_{\pm}
\eqno (100)
$$
The relations (99) and (100) are basic ingredients for 
the definition of the Hopf algebra $U_q(sl(2))$. 

\sb 
\sd

\noindent {\bf APPENDIX}

\sb

Let $Q \in {\bf C} \setminus \{ 1 \}$. The application 
$$
[~~]_Q : {\bf R} \to {\bf C} : x \mapsto [x]_Q := { {1 - Q^x} \over {1 - Q} }
\eqno ({\rm A.}1)
$$
generates $Q$-deformed numbers $[x]_Q$. We have
$$
[n]_Q = \sum_{i = 0}^{n - 1} Q^i \quad n \in {\bf N} \setminus \{ 0 \} 
\eqno ({\rm A.}2)
$$
when $x = n$ is a strictly positive integer. In the case where 
$Q = q$ ($q := {\rm exp} ( 2 \pi {\rm i} / k)$ being the root of unity, 
with $k \in {\bf N} \setminus \{ 0,1 \}$, considered in the main body 
of the present paper), we have   
$$
[x]_q = {\rm exp} \left[ (x - 1) {\pi \over k} {\rm i} \right]
{ {\sin {\pi \over k} x } \over 
  {\sin {\pi \over k}   } }
\eqno ({\rm A.}3) 
$$
for any real number $x$. In this case, it is clear that
$[x]_{\bar q}$ is the complex conjugate of 
$[x]_{     q}$.

\sb 
\sd

\noindent
{\bf Acknowledgments}

\sb 

One of the authors (M.K.) is very indebted to B.~Gruber 
to have given him the opportunity to present this work 
to the beautiful symposium ``Symmetries in Science X''. 
He is also grateful to W.S.~Chung, G.A.~Goldin, B.~Gruber,
M.~Lorente, V.I.~Man'ko, S.~Mashkevich and M.~Moshinsky 
for interesting comments on this work.

\sb
\sd

\noindent {\bf REFERENCES}

\sb

\item{1.} 
  V.A. Rubakov and V.P. Spiridonov, 
  {\it Mod. Phys. Lett. A} {\bf 3} (1988) 1337. 

\item{2.}  
 A.T. Filippov, A.P. Isaev and A.B. Kurdikov, {\it Mod. Phys. Lett. A} {\bf 7} 
 (1992) 2129~; ibid., {\it Int. J. Mod. Phys. A} {\bf 8} (1993) 4973. 

\item{3.}
  A. Le Clair and C. Vafa, {\it Nucl. Phys. B} {\bf 401} (1993) 413. 

\item{4.}  
 S. Durand, {\it Phys. Lett. B} {\bf 312} (1993) 115. 

\item{5.}  
 N. Debergh, {\it J. Phys. A} {\bf 26} (1993) 7219. 

\item{6.}  
 N. Mohammadi, preprint hep-th/9412133. 

\item{7.}  
 J.L. Matheus-Valle and M.A. R.-Monteiro, {\it Phys. Lett. B}
 {\bf 300} (1993) 66. 

\item{8.}  
 N. Fleury and M. Rausch de Traubenberg, {\it Mod. Phys. Lett. A} 
 {\bf 27} (1996) 899. 

\item{9.}  
 R. Kerner, {\it J. Math. Phys.} {\bf 33} (1992) 403. 

\item{10.}  
 J. Beckers and N. Debergh, {\it Mod. Phys. Lett. A} {\bf 4} (1989) 1209. 

\item{11.}  
 J.A. de Azc\'arraga and A. J. Macfarlane, 
 {\it J. Math. Phys.} {\bf 37} (1996) 1115. 

\item{12.}  
 M. Arik and D.D. Coon, {\it J. Math. Phys.} {\bf 17} (1976) 524. 

\item{13.}  
  A.J. Macfarlane, {\it J. Phys. A} {\bf 22} (1989) 4581. 

\item{14.}  
 L.C. Biedenharn, {\it J. Phys. A} {\bf 22} (1989) L873. 

\item{15.}  
 M. Chaichian, D. Ellinas and P.P. Kulish, 
 {\it Phys. Rev. Lett.} {\bf 65} (1990) 980. 

\item{16.}  
 J. Katriel and A.I. Solomon, {\it J. Phys. A} {\bf 24} (1991) 2093. 
  
\item{17.}  
 R.J. McDermott and A.I. Solomon, {\it J. Phys. A} {\bf 27} (1994) L15. 
  
\item{18.}  
 R.J. McDermott and A.I. Solomon, {\it J. Phys. A} {\bf 27} (1994) 2037. 
  
\item{19.}  
 V.I. Man'ko, G. Marmo, E.C.G. Sudarshan and F. Zaccaria, 
 {\it Phys. Scripta} {\bf 55} (1997) 520. 
  
\item{20.}  
 T.K. Kar and Gautam Ghosh, {\it J. Phys. A} {\bf 29} (1996) 125. 
   
\item{21.}  
 S. Majid and M.J. Rodr\'\i guez-Plaza, 
 {\it J. Math. Phys.} {\bf 35} (1994) 3753. 
   
\item{22.}  
 A.M. Perelomov, {\it Generalized Coherent States and Their Applications} 
 (Springer, Berlin, 1986). 

\item{23.}  
 R.J. Glauber, {\it Phys. Rev.} {\bf 131} (1963) 2766. 

\item{24.}  
 E.C.G. Sudarshan, {\it Phys. Rev. Lett.} {\bf 10} (1963) 84. 

\item{25.}  
 M.M. Nieto and L.M. Simmons, Jr., {\it Phys. Rev. Lett.} {\bf 41} (1978) 207. 

\item{26.}  
 R.S. Dunne, A.J. Macfarlane, J.A. de Azc\'arraga and J.C. P\'erez Bueno,
 {\it Phys. Lett. B} {\bf 387} (1996) 294. 

\item{27.}  
 R.S. Dunne, A.J. Macfarlane, J.A. de Azc\'arraga and J.C. P\'erez Bueno,
 preprint hep-th/96100087. 

\item{28.}  
 Y. B\'erub\'e-Lausi\`ere and V. Hussin, {\it J. Phys. A} {\bf 26} (1993) 6271. 

\item{29.}  
 M. Daoud, Y. Hassouni and M. Kibler, work in progress. 
  
\item{30.}  
 P.A.M. Dirac, {\it Proc. R. Soc.} {\bf 114} (1927) 243.  
  
\item{31.}  
 W.P. Schleich and S.M. Barnett, {\it Phys. Scripta} {\bf 48} (1993) 243.  

\item{32.}  
 D.T. Pegg and S.M. Barnett, {\it Europhys. Lett.} {\bf 6} (1988) 6~; 
 ibid., {\it Phys. Rev. A} {\bf 39} (1989) 1665.

\item{33.}  
 D. Ellinas, {\it Phys. Rev. A} {\bf 45} (1992) 3358. 
  
\item{34.}  
  D.F. Fairlie, P. Fletcher and C.K. Zachos, {\it J. Math. Phys.} 
  {\bf 31} (1990) 1088. 
  
\item{35.}  
 H.T. Sato, {\it Mod. Phys. Lett. A} {\bf 9} (1994) 1819~; 
 ibid., {\it Prog. Theo. Phys.} {\bf 93} (1994) 195. 
  
\item{36.}  
 I.I. Kogan, {\it Mod. Phys. Lett. A} {\bf 7} (1992) 3717~; 
 ibid., {\it Int. J. Mod. Phys. A} {\bf 9} (1994) 3889. 
  
\bye